\documentclass[aps,prl,twocolumn]{revtex4}
\usepackage{graphicx}
\usepackage{amsmath,amssymb}
\usepackage{bm}
\usepackage{verbatim}
\usepackage{ulem}
\begin{document}
\title{Probing the topological exciton condensate via Coulomb drag}
\author{M. P. Mink}
\email{m.p.mink@uu.nl}
\affiliation{Institute for Theoretical Physics, Utrecht
University, Leuvenlaan 4, 3584 CE Utrecht, The Netherlands}
\author{H. T. C. Stoof}
\affiliation{Institute for Theoretical Physics, Utrecht
University, Leuvenlaan 4, 3584 CE Utrecht, The Netherlands}
\author{Marco Polini}
\affiliation{NEST, Istituto Nanoscienze-CNR and Scuola Normale Superiore, I-56126 Pisa, Italy}
\author{G. Vignale}
\affiliation{Department of Physics and Astronomy, University of Missouri, Columbia, Missouri 65211, USA}
\author{R. A. Duine}
\affiliation{Institute for Theoretical Physics, Utrecht
University, Leuvenlaan 4, 3584 CE Utrecht, The Netherlands}
\begin{abstract}
The onset of exciton condensation in a topological insulator thin film was recently predicted. We calculate the critical temperature for this transition, taking into account screening effects. Furthermore, we show that the proximity to this transition can be probed by measuring the Coulomb drag resistivity between the surfaces of the thin film as a function of temperature. This resistivity shows an upturn upon approaching the exciton-condensed state.
\end{abstract}
\maketitle

{\it Introduction. ---} Recently, there has been great experimental and theoretical interest in a new class of materials called topological insulators (TIs)~\cite{hasan}. These materials combine, for a particular doping level, an insulating behavior in the bulk with topologically protected conducting surface states that are described by a two-dimensional (2D) Dirac-Weyl Hamiltonian. When the surfaces of a TI film are independently doped or gated, electrons can be induced in one layer and holes in the other. In 2009, Seradjeh {\it et al.}~\cite{franz} predicted that such a system can support a so-called topological exciton condensate. Electrons in one layer combine with holes in the other layer to form excitons, which condense for low enough temperatures. One of the interesting features of this state is that it supports vortices in the order parameter: these vortices  carry a fractional charge $\pm e/2$~\cite{franz}.
\begin{center}
\begin{figure}
\includegraphics[width = 1.0 \linewidth]{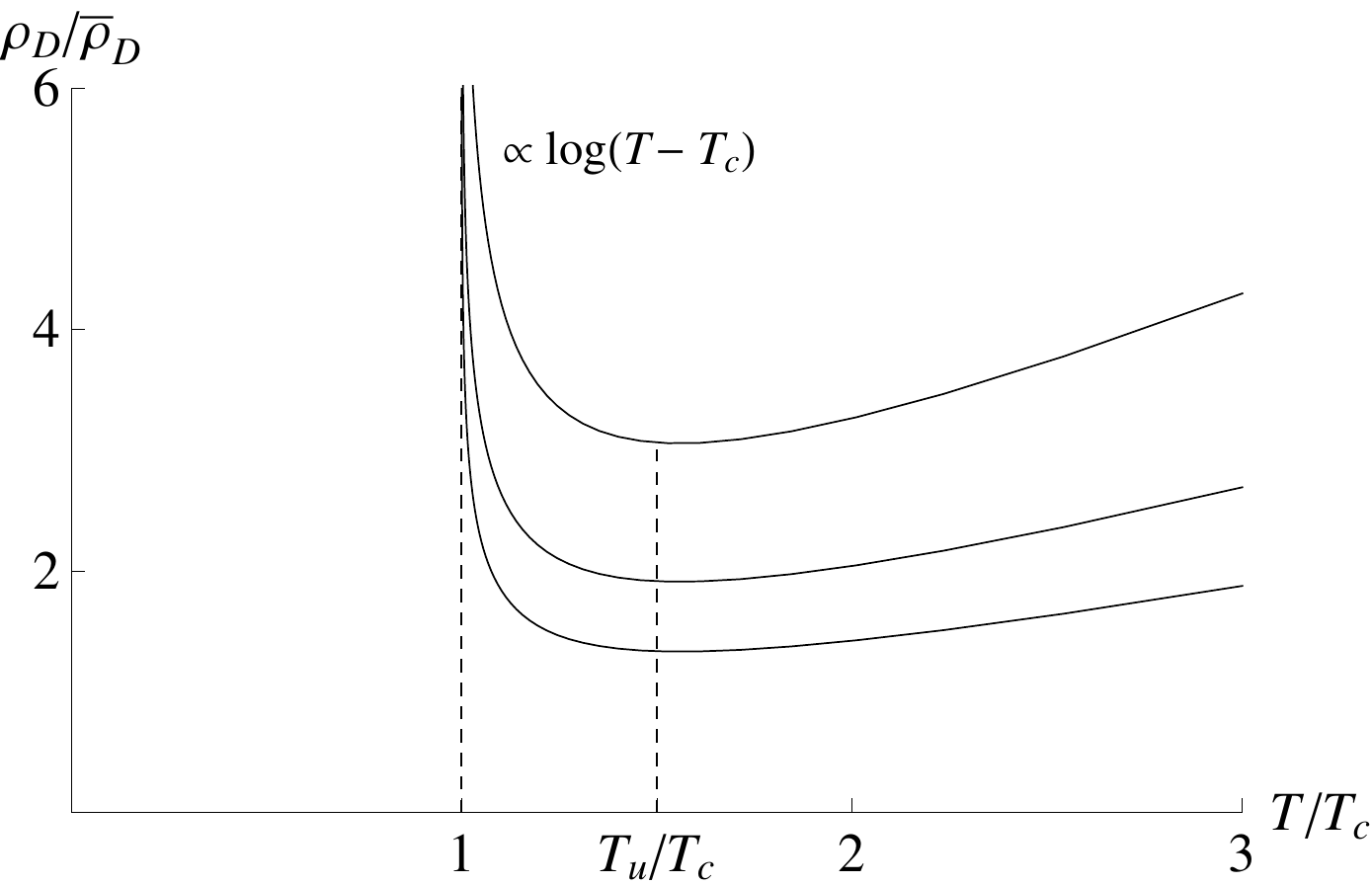}
\caption{The drag resistivity $\rho_{\rm D}$ [in units of
${\bar \rho}_{\rm D} = (\nu_0 U)^2 \hbar/e^2$, with $\nu_0$ the density of states at the Fermi level and $U$ the interlayer interaction strength] calculated from Eq.~(\ref{eq:sd}) as a function of the reduced temperature $T/T_{\rm c}$. With this choice of units $\rho_{\rm D}/{\bar \rho}_{\rm D}$ depends on two parameters -- the chemical potential $\mu$ and the critical temperature $T_{\rm c}$ -- which implicitly determine the interaction strength $U$ in Eq.~(\ref{eq:Vegen}). The curves correspond to the case of equal chemical potentials in the two layers ($\mu = 0.04~{\rm eV}$). From top to bottom the critical temperatures are $T_{\rm c}/T_{\rm F} = (1/150,1/200,1/250)$. The vertical dashed lines are located at $T = T_{\rm c}$ (left) and at the ``upturn" temperature scale (see text) $T = T_{\rm u}$ (right). \label{fig:one}}
\end{figure}
\end{center}
In a drag experiment with a two-layer system a current is applied through one of the layers (the drive or active layer) and an induced voltage is measured in the other layer (the passive layer). The drag resistivity $\rho_{\rm D}$ is defined as the ratio between the electric field in the passive layer and the current density in the active layer. In the absence of an ordered state, the low-temperature behavior of the drag resistivity is $\rho_{\rm D} \propto T^2$ [or $T^2 \log(T)$ for a 2D system with short-range interactions]. This quadratic temperature dependence was first observed in 1991 by Gramila {\it et al.}~\cite{gramila} and is a hall mark of Fermi-liquid behavior. A review of drag effects in two-layer systems is given in Ref.~\cite{rojo}. A departure from the standard $\rho_{\rm D} \propto T^2$ temperature dependence indicates the occurrence of non-Fermi-liquid behavior. The low-temperature dependence of $\rho_{\rm D}$ for semiconductor electron-hole bilayers was studied experimentally in numerous recent works~\cite{pepper,lilly}. On the theoretical side, we note the calculation by Hu~\cite{hu}, which predicted an enhancement of Coulomb drag for  ordinary electron-hole bilayers at temperatures just above the excitonic instability, and the analysis of Ref.~\cite{finsigma}, which predicted a sharp increase of the drag resistivity as the temperature is lowered below the critical temperature of condensation.

In this Letter we show that the drag resistivity shows a precursor of the topological exciton condensed phase, {\it via} an upturn upon approaching the critical temperature, as shown in Fig.~\ref{fig:one}. Close to the critical temperature we find that on the basis of a Boltzmann analysis $\rho_{\rm D} \propto \log(T-T_{\rm c})$. This upturn could therefore be used to determine the proximity to the phase transition. We note that the system considered here resembles double-layer graphene (DLG) --  a system of two graphene layers separated by a dielectric barrier. The main differences are the number of degenerate electron species and the dielectric constant~\cite{lozovik,tutuc}. The results presented here are therefore also qualitatively applicable to DLG, a system for which exciton condensation has been predicted too~\cite{lozovik,min,joglekar}. Coulomb drag in DLG has been intensively studied theoretically~\cite{CDnormal}. This literature, however, refers to drag in which both layers are either electron- or hole-doped so that exciton condensation does not occur.

{\it Coulomb drag and exciton condensation. ---} We consider a  TI thin film whose top layer is hole-doped and whose bottom layer is electron-doped. We apply the ``closed-band approximation"~\cite{lozovik,min,joglekar}, {\it i.e.}, we consider only the upper Dirac cone for the electron layer and the lower Dirac cone for the hole layer.  This is  justified since the temperatures we consider are sufficiently small and the screening lengths sufficiently large, so that the far-lying bands can in first instance be considered inert. Then, the dispersions are well approximated by $\epsilon_{\rm b}({\bm k}) = v |{\bm k}| - \mu_{\rm b}$ and $\epsilon_{\rm t}({\bm k}) = - v |{\bm k}| + \mu_{\rm t}$ for the bottom and top layer, respectively ($\hbar =1$ throughout this Letter). Here $v \approx 5\times 10^{5}~{\rm m}/{\rm s}$ is the Fermi velocity appropriate for the surface states of a typical TI thin film~\cite{bandvelo}.  These formulas are valid up to a cutoff $\xi = 0.2~{\rm eV}$, which is the typical distance between the Dirac point and the bulk bands in a TI. We also introduce the mean Fermi energy, $\mu = (\mu_{\rm b} + \mu_{\rm t})/2$, and half the chemical potential imbalance, $h = (\mu_{\rm b}-\mu_{\rm t})/2$.

\begin{center}
\begin{figure}
\includegraphics[width = 1.0 \linewidth]{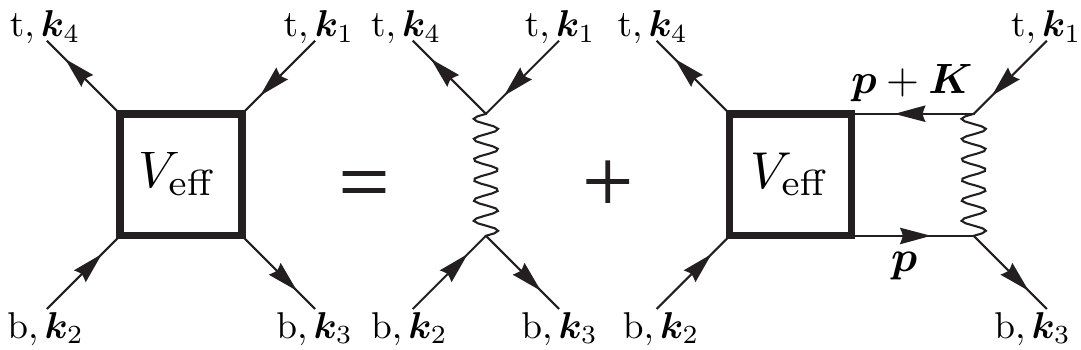}
\caption{The Bethe-Salpeter equation for the effective interaction $V_{\rm eff}$ -- Eq.~(\ref{eq:Vegen}). The thin wavy line is $V_0$ and the arrows are the electron propagators in the top layer (top arrows) and the bottom layer (bottom arrows). Furthermore, ${\bm K} = {\bm k}_1 - {\bm k}_3 = {\bm k}_4 - {\bm k}_2$  is the center-of-mass momentum of the electron-hole pair and  ${\bm q} = {\bm k}_1 - {\bm k}_4 = {\bm k}_3 - {\bm k}_2$ is the interlayer momentum transfer, which is responsible for  Coulomb drag.  Finally, ${\bm p}$ is an integration variable.  To make contact with the drag resistivity in Eq.~(\ref{eq:sd}) replace  ${\bm k} = {\bm k}_1$ and ${\bm k}' = {\bm k}_2$.\label{fig:two}}
\end{figure}
\end{center}

Our theoretical treatment is based on the  Boltzmann equation, whose main merit is to give a physically transparent picture of the scattering processes that control the  momentum transfer rate between the layers. The crucial quantity in the Boltzmann approach is the scattering amplitude $V_{\rm eff}({\bm k}_1,{\bm k}_2,{\bm k}_3,{\bm k}_4)$, which is diagrammatically presented on the left-hand side in Fig.~\ref{fig:two}, where
${\bm k}_1$ and ${\bm k}_2$ are incoming momenta, one from each layer, and ${\bm k}_3$ and ${\bm k}_4$ outgoing momenta.
 
Under ordinary circumstances this interaction is well approximated by the single screened interaction line $V_0({\bm k}_1 - {\bm k}_4)$, which is shown as a wavy line in Fig.~\ref{fig:two}. But this simple picture breaks down in the vicinity of the exciton pairing transition, where it becomes necessary to include the effect of pairing fluctuations, diagrammatically represented by the infinite series of ladder diagrams shown in the right-hand side Fig.~\ref{fig:two}.  This series of diagrams diverges at the critical temperature  $T_{\rm c}$ when the center-of-mass momentum of the electron-hole pair ${\bm K} = {\bm k}_1 - {\bm k}_3$ and the corresponding energy $\Omega=\epsilon_{\rm b}({\bm k}_1)-\epsilon_{\rm t}({\bm k}_3)$  tend to zero. The series can be calculated analytically if the momentum dependence of  the screened interaction $V_0$ is neglected, which is  a reasonable approximation in most cases and sufficient for a qualitative understanding of the effect of pairing on Coulomb drag. One then obtains an effective interaction that depends only on the center-of-mass momentum ${\bm K}$ and energy $\Omega$ as
\begin{equation} \label{eq:Vegen}
V({\bm k}_1,{\bm k}_2,{\bm k}_3,{\bm k}_4) \simeq V_{\rm eff}({\bm K},\Omega) \equiv  \frac{U}{1 - U \Xi({\bm K},\Omega)}~,
\end{equation}
where $U$ is the momentum-independent contact interaction strength representing the screened interaction and
\begin{equation} \label{eq:Pgen}
\Xi({\bm K},\Omega)= \frac{1}{A} \sum_{\bm k} \frac{n(\epsilon_{\rm t}({\bm k} + {\bm K})) - n(\epsilon_{\rm b}({\bm k}))}{\epsilon_{\rm b}({\bm k}) - \epsilon_{\rm t}({\bm k} + {\bm K}) - \Omega -i 0^+}
\end{equation}
is a pairing susceptibility, $A$ being the 2D electron system area. The temperature $T$ enters the above expression through the Fermi-Dirac distribution, $n(E) = 1/[1+\exp(\beta E)]$, where $\beta = (k_{\rm B} T)^{-1}$.
It is easy to check that, in the symmetric case ($h=0$) $\Xi({\bm 0},0)$ diverges logarithmically at zero temperature, and therefore the denominator of Eq.~(\ref{eq:Vegen}) must vanish  at some non-zero temperature $T_{\rm c}$ no matter how small the electron-hole attraction.  This is the mean-field pairing transition temperature.   We see that, as the temperature is decreased towards $T_{\rm c}$, pairing fluctuations first lead to an enhancement of $V_{\rm eff}$ for small ${\bm K}$ and $\Omega$ and ultimately to a pole at $\Omega = 0$ and ${\bm K} = {\bm 0}$ as $T = T_{\rm c}$. The occurrence of this pole is the cause for the upturn of the Coulomb drag resistivity close to $T_{\rm c}$.

Inserting Eq.~(\ref{eq:Vegen}) in the  Boltzmann collision integral and performing standard manipulations, we finally arrive at the following expression for the drag resistivity $\rho_{\rm D}$:
\begin{multline}\label{eq:sd}
\rho_{\rm D} =
-\frac{\beta}{2 (2 \pi)^6 e^2 n v^2} \int d {\bm K}  d\Omega
\frac{|V_{\rm eff}({\bm K},\Omega)|^2}{\sinh^2(\beta\Omega/2)}  \\
\times \int d {\bm k} d{\bm k}' \Im m\left[b({\bm k}; {\bm K},\Omega)\right] \Im m\left[b({\bm k}'; {\bm K},\Omega)\right] \\
\times [{\bm v}_{\rm t}({\bm k}' + {\bm K}) - {\bm v}_{\rm t}({\bm k} + {\bm K})]
\cdot [{\bm v}_{\rm b}({\bm k}')  -   {\bm v}_{\rm b}({\bm k})]~,
\end{multline}
where ${\bm v}_{{\rm t}({\rm b})}({\bm k}) \equiv {\bm \nabla} \epsilon_{{\rm t}({\rm b})}({\bm k})$ are the group velocities, $n$ is the single layer carrier density, and
\begin{equation}
b({\bm k}; {\bm K}, \Omega) \equiv \frac{n(\epsilon_{\rm t}({\bm k} + {\bm K})) - n(\epsilon_{\rm b}({\bm k}))}{\epsilon_{\rm b}({\bm k}) - \epsilon_{\rm t}({\bm k} + {\bm K}) - \Omega -i 0^+}
\end{equation}
is the summand of  Eq.~(\ref{eq:Pgen}). Since ${\bm v}_{\rm t}({\bm k})$ and ${\bm v}_{\rm b}({\bm k})$  are oppositely directed $\rho_{\rm D}$ is positive, as expected for carriers of opposite polarities.

{\it Drag Resistivity. ---} To determine the drag resistivity we need to evaluate Eq.~(\ref{eq:sd}) numerically.
However, the qualitative behavior near $T_{\rm c}$ can be obtained analytically as follows (in what follows we first consider the balanced case $h =0$).
First, we notice that the denominator of Eq.~(\ref{eq:Vegen}) can be expanded, for small $K$ and $\Omega$ as follows:
\begin{eqnarray}\label{eq:smallKOmega}
1 - U \Xi({\bm K},\Omega) &\simeq& \alpha(T) + a(T) U \nu_0 (\beta v K)^2 \nonumber\\
&+& i U \nu_0 \beta \Omega/4~,
\end{eqnarray}
where $\alpha(T) \propto T-T_{\rm c}$ measures the distance from the critical point,
$\nu_0 = \mu/(2\pi v^2)$ is the density-of-states at the Fermi energy, and $a(T)$ is a (dimensionless) positive ultraviolet-convergent quantity. Further, for ${\bm K} \to {\bm 0}$ we have $[{\bm v}_{\rm t}({\bm k}' + {\bm K}) - {\bm v}_{\rm t}({\bm k} + {\bm K})]
\cdot [{\bm v}_{\rm b}({\bm k}')  -   {\bm v}_{\rm b}({\bm k}) ]=  2v^2 [\cos(\phi) -1]$, where $\phi$ is the angle between ${\bm k}$ and ${\bm k}'$. Since this quantity vanishes only in a set of zero measure with respect to the two 2D integrals over ${\bm k}$ and ${\bm k}'$ in Eq.~(\ref{eq:sd}), we can approximate the scalar product
with a momentum-independent quantity of the order of $v^2$. The integrals over ${\bm k}$ and ${\bm k}'$ can now be carried out analytically: using that
$\int d {\bm k}~\Im m[b({\bm k}; {\bm 0},\Omega)] = \Im m[\Xi({\bm 0},\Omega)] = - \nu_0\beta \Omega/4$
for $\beta \Omega \ll 1$, we obtain
\begin{eqnarray} \label{eq:sdtc}
\rho_{\rm D} &\propto&
 \int dK  d\Omega \frac{(\beta \Omega/4)^2}{\sinh^2(\beta \Omega/2)} \nonumber \\
&\times& \frac{(\nu_0U)^2 K}{\left[\alpha(T)+a U\nu_0(\beta v K)^2\right]^2+ (U \nu_0\beta\Omega/4)^2}~,
\end{eqnarray}
which is immediately seen to diverge logarithmically when $T\to T_{\rm c}$ ({\it i.e.}, for $\alpha \to 0$).

In Fig.~\ref{fig:one} we show the numerically evaluated drag resistivity $\rho_{\rm D}$ as a function of temperature. We observe that for temperatures much larger than $T_{\rm c}$, the drag resistivity increases quadratically as $\rho_{\rm D} \propto T^2$~\cite{log}. The logarithmic divergence of $\rho_{\rm D} \propto \log(T-T_{\rm c})$ is clearly visible. From top to bottom the curves correspond to $T_{\rm c}/T_{\rm F} = (1/150,1/200,1/250)$. Following Ref.~\cite{lilly}, we define $T_{\rm u}$ as the temperature at which the upturn of $\rho_{\rm D}$ starts, as shown in Fig.~\ref{fig:one}. As a rule of thumb, $T_{\rm u} \approx (3/2) T_{\rm c}$.

{\it Critical temperature. ---} In order to assess whether the upturn in the drag resistivity can be observed experimentally, we need to calculate the (mean-field~\cite{BKT}) critical temperature as a function of interlayer distance and carrier density, taking into account screening effects~\cite{kharitonov}. We employ a separable approximation to the momentum-dependent interlayer interaction $V_0({\bm k}_1 - {\bm k}_4)$~\cite{separable}. The advantage of this approach over using a contact interaction is that it captures the decrease of the interaction strength $V_0$ with increasing transferred momenta. Furthermore, no ultraviolet cutoff on which the result for $T_{\rm c}$ would depend is needed.

The screened interlayer Coulomb interaction is given by $V_0(q, \omega) = V_{\rm tb}(q)/\varepsilon(q,\omega)$, where $V_{\rm tb}(q)$ is the bare interaction between one electron on the top surface and one on the bottom surface, $q = |{\bm k} - {\bm k}'|$ and $\omega = \epsilon_{\rm b}({\bm k}) - \epsilon_{\rm b}({\bm k}')$ are momentum and energy transfers, respectively, and  $\varepsilon(q,\omega)$ is the  dielectric screening function of the carriers,  which we approximate in the random phase approximation. Both  $V_{\rm tb}(q)$ and $\varepsilon(q,\omega)$ depend on the dielectric constants of the TI and of the environment surrounding the thin film. The details of the construction are given in Ref.~\cite{profumo}.

\begin{center}
\begin{figure}
\includegraphics[width = 1.0 \linewidth]{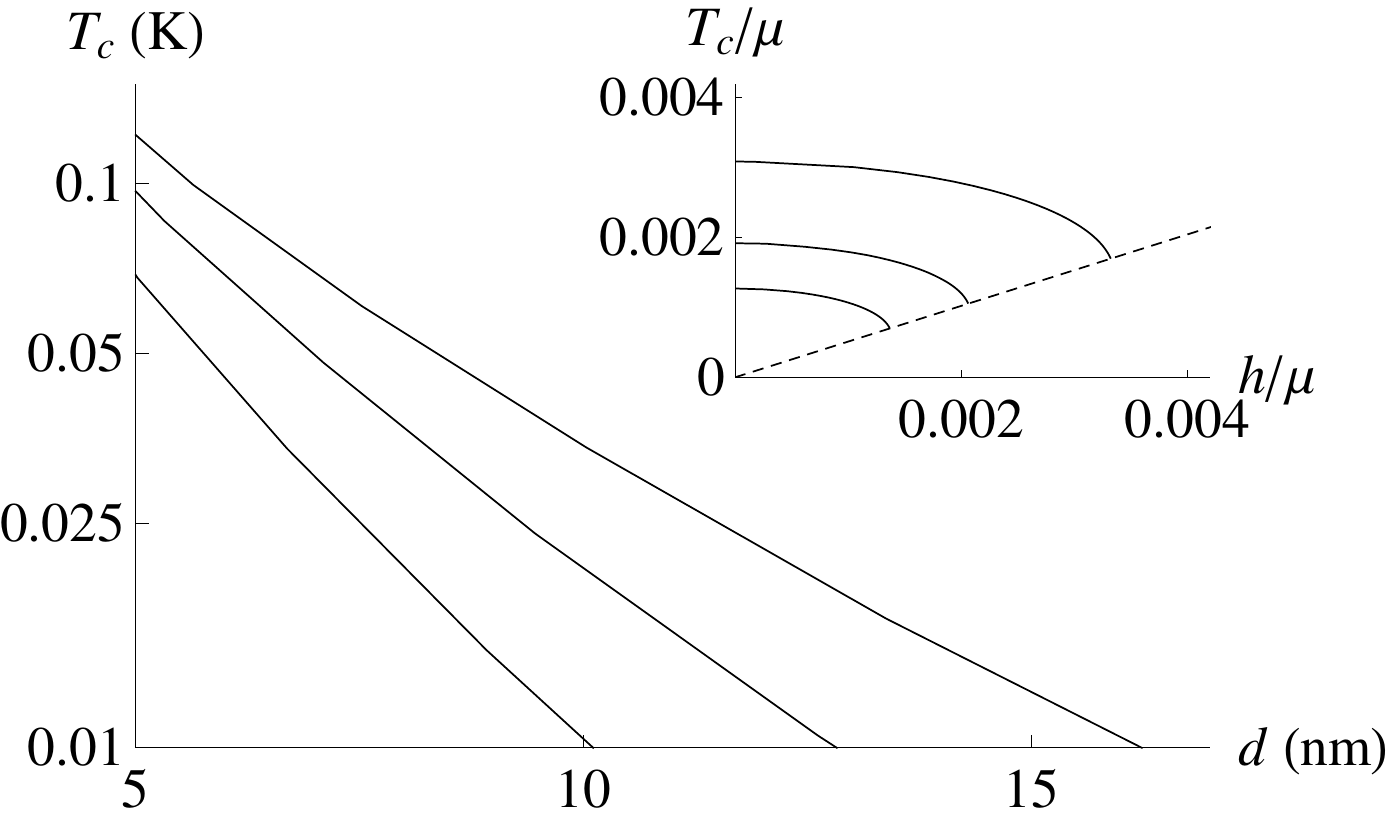}
\caption{Critical temperature $T_{\rm c}$ (in ${\rm K}$) versus interlayer distance $d$ (in ${\rm nm}$) in the absence of density imbalance ($h=0$). From right to left the lines correspond to carrier densities $n= (0.25, 0.5, 1) \times 10^{10}~{\rm cm}^{-2}$. Inset: The critical lines $T_{\rm c} /\mu$ versus $h/\mu$. From top to bottom the lines correspond to $k_{\rm F} d = (0.05,0.075,0.1)$. The critical lines terminate on the dashed line, {\it i.e.},  the locus of the points where  the fourth order term of the Ginzburg-Landau expansion of the free energy in powers of the order parameter vanishes.\label{fig:three}}
\end{figure}
\end{center}

We consider a TI film with vacuum above the top surface and a typical substrate material ({\it e.g.} SiO$_2$) below the bottom surface: the appropriate background dielectric constants are $\epsilon_{\rm top} = 1,~\epsilon_{\rm TI} =100,~\epsilon_{\rm bottom}=4$. For the balanced case with $h=0$ we obtain the phase diagram in Fig.~\ref{fig:three}. From right to left the lines correspond to carrier densities $n= (0.25,0.5,1) \times
10^{10}~{\rm cm}^{-2}$. Our results show that the critical temperatures are within reach using existing cryogenic techniques for solid-state systems. Furthermore, also the required surface carrier densities have been reached previously \cite{hasan}. In the inset of Fig.~\ref{fig:three} we show the effect of a non-zero density imbalance. The solid curves represent the behavior of $T_{\rm c}/\mu$ for constant $k_{\rm F} d$. From top to bottom the curves correspond to $k_{\rm F} d = (0.05,0.075,0.1)$. We note that $T_{\rm c}/\mu$ decreases strongly with increasing density imbalance. The dashed line is the line along which the quartic term of the Landau-Ginzburg expansion of the free energy in powers of the order parameter, evaluated at $T_{\rm c}$, vanishes.  Thus, for a fixed $k_{\rm F} d$, the transition becomes first order as $h$ grows beyond the intersection of the dashed line and the curve corresponding to that $k_{\rm F} d$ value. The first-order region below the dashed line cannot be accurately described by our present normal-state formalism.
The decreasing behavior of $T_{\rm c}$ with $h$ indicates that the excitonic transition is strongly suppressed by the presence of any density imbalance. Thus, it is important in experimental realizations to have accurate control over the electron and hole densities in the two layers.

{\it Discussion and conclusions. ---} Our theory predicts a divergence of $\rho_{\rm D}$ as $T$ approaches $T_{\rm c}$, but does not include the possible suppression of this divergence by critical fluctuations in the immediate vicinity of $T_{\rm c}$.
We expect that inclusion of these fluctuations leads to a down turn of $\rho_{\rm D}$ for temperatures very close to the critical temperature~\cite{critfluc}, before  increasing again upon further decrease of the temperature, as predicted in Ref.~\cite{finsigma}.

We note that if the layers are brought too close together, the wave functions in the top layer and the bottom layer begin to overlap and interlayer tunneling becomes important. This limits our theory to a minimal interlayer distance $d \gtrsim 10~{\rm nm}$. Actually, also conduction over the side layers of the TI film could be important. One expects that for samples of large enough area, these can be neglected. If not, a transport-prohibiting gap could be induced in the side layers by interfacing those with a ferromagnet.

In closing, we note that the upturn of $\rho_{\rm D}$  is analogous to the upturn of the spin-drag resistivity recently predicted in a cold gas of fermionic atoms near a ferromagnetic transition~\cite{duine_prl_2010}. In particular, spin-fluctuation contributions to the effective interaction in the ferromagnetic case play the same role as pairing fluctuations in the present case. The reason why in Ref.~\cite{duine_prl_2010} the spin drag resistivity was predicted to remain finite at $T_{\rm c}$ is that the transverse spin-fluctuation channel was treated as part of the familiar (Hartree) charge and longitudinal spin-density fluctuation channels. When transverse spin fluctuations in the Fock channel are also included, a logarithmic divergence is obtained, just as in the present case. We also note that our results, while qualitatively similar to those obtained in Ref.~\cite{hu} for massive electrons and holes in a conventional electron-hole semiconductor bilayer, differ, at the theoretical level, by the inclusion of an additional term in the collision integral. This term is easily missed by using the Kubo formula. The nature of the additional term is similar to the vertex corrections that are responsible for replacing the momentum lifetime by the transport lifetime in the classical Drude formula for the resistivity.

\acknowledgments
This work was supported by the Stichting voor Fundamenteel Onderzoek der Materie (FOM), the Netherlands Organization for Scientific Research (NWO), and by the European Research Council (ERC). G.V. was supported by the US Department of Energy grant DE-FG02-05ER46203.


\begin{thebibliography}{77}
%
\bibitem{hasan}
M.Z. Hasan and C.L. Kane, \rmp {\bf 82}, 3045 (2010);
X.-L. Qi and S.-C. Zhang, arXiv:1008.2026.
%
\bibitem{franz}
B. Seradjeh, J.E. Moore, and M. Franz, \prl {\bf 103}, 066402 (2009); 
see also Z. Wang {\it et al.}, arXiv:1106.5838.
%
\bibitem{gramila}
T.J. Gramila {\it et al.}, \prl  {\bf 66}, 1216 (1991).
%
\bibitem{rojo}
A.G. Rojo,  J. Phys.: Condens. Matter {\bf 11}, R31 (1999).
%
\bibitem{pepper}
A.F. Croxall {\it et al.}, \prl {\bf 101}, 246801 (2008).
%
\bibitem{lilly}
J.A. Seamons {\it et al.}, \prl {\bf 102}, 026804 (2009).
%
\bibitem{hu}
B. Y.-K. Hu, \prl {\bf 85}, 820 (2000).
%
\bibitem{finsigma}
G. Vignale and A.H. MacDonald, \prl {\bf 76}, 27856 (1996).
%
\bibitem{lozovik}
Yu. E. Lozovik and A.A. Sokolik, JETP Lett. {\bf 87}, {\bf 55} (2008).
%
\bibitem{tutuc}
S. Kim {\it et al.}, \prb {\bf 83}, 161401 (2011).
%
\bibitem{min}
H. Min {\it et al.}, \prb {\bf 78}, 121401(R) (2008).
%
\bibitem{joglekar}
C.-H. Zhang and Y.N. Joglekar,  \prb {\bf 77}, 205426 (2008).
%
\bibitem{CDnormal}
W.-K. Tse, B. Y.-K. Hu, and S. Das Sarma, \prb {\bf 76}, 081401(R) (2007);
N.M.R. Peres, J.M.B. Lopes dos Santos, and A.H. Castro Neto, Europhys. Lett. {\bf 95}, 18001 (2011);
M.I. Katsnelson, \prb {\bf 84}, 041407(R) (2011);
E.H. Hwang and S. Das Sarma, arXiv:1105.3203v1.
%
\bibitem{bandvelo}
See {\it e.g.} Y. Xia {\it et al.}, Nature Phys. {\bf 5}, 398 (2009).
%
\bibitem{log}
In the absence of pairing fluctuations, the dominant low-temperature behavior is $\rho_{\rm D} \propto T^2 \log(T)$ and not $\rho_{\rm D} \propto T^2$. The logarithmic correction is an artifact of the contact interaction used.
%
\bibitem{BKT}
Berezinskii-Kosterlitz-Thouless (BKT) phase fluctuations tend to suppress $T_{\rm c}$ with respect to our mean-field estimate. The BKT temperature is given by $T_{\rm BKT} = \pi \rho_{\rm s}(T_{\rm BKT})/2$, where $\rho_{\rm s}(T)$ is the superfluid density, which can be estimated from a mean-field analysis. For DLG exciton condensates this analysis has been carefully carried out in Ref.~\cite{min}.
%
\bibitem{kharitonov}
M. Yu. Kharitonov and K.B. Efetov, \prb {\bf 78}, 241401 (2008); see also R. Bistritzer {\it et al.}, arXiv:0810.0331.
%
\bibitem{separable}
The $T_{\rm c}$-equation is $1 =  (1/A)\sum_{\bm k} V^2_{\rm sep}(k) [n(\epsilon_{\rm t}({\bm k})) - n(\epsilon_{\rm b}({\bm k}))]/[\epsilon_{\rm b}({\bm k}) - \epsilon_{\rm t}({\bm k})]$, where $V_{\rm sep}(k)$ is a separable approximation to the angular average of $V_0$ over incoming and outgoing momenta, {\it i.e.} $V_{\rm sep}(k) = V_{\rm av}(k,k_{\rm F})/V_{\rm av}(k_{\rm F}, k_{\rm F})$
with $V_{\rm av}(k,k') = \int d \phi \left.V_0(q)\right|_{q = |{\bm k} - {\bm k}'|}[1 + \cos(\phi)]/(4 \pi)$,
where we have included the form factor $[1 + \cos(\phi)]/2$ relevant for this system [Yu. E. Lozovik and A.A. Sokolik, Eur. Phys. J. B {\bf 25}, 195 (2010); M.P. Mink {\it et al.}, arXiv:1107.4477].
%
\bibitem{profumo}
R.E.V. Profumo {\it et al.}, \prb {\bf 82}, 085443 (2010).
%
\bibitem{critfluc}
R. Kittinaradorn, R.A. Duine, and H.T.C. Stoof, arXiv:1107.2024.
%
\bibitem{duine_prl_2010}
R.A. Duine {\it et al.}, \prl {\bf 104}, 220403 (2010).
%
\end{thebibliography}
\end{document}